\documentclass[12pt]{iopart}
\usepackage{pstricks,pst-node,pst-slpe}

\eqnobysec

\newcommand{\beq}{\begin{equation}}
\newcommand{\eeq}{\end{equation}}
\newcommand{\ba}[1]{\begin{array}{#1}}
\newcommand{\ea}{\end{array}}
\newcommand{\bea}{\begin{eqnarray}}
\newcommand{\eea}{\end{eqnarray}}
\newcommand{\nn}{\nonumber \\}
\newcommand{\ben}{\begin{enumerate}}
\newcommand{\een}{\end{enumerate}}
\newcommand{\bit}{\begin{itemize}}
\newcommand{\eit}{\end{itemize}}
\newcommand{\bde}{\begin{description}}
\newcommand{\ede}{\end{description}}

\newcommand{\ds}{\displaystyle}

\newcommand{\sz}{\scriptsize}

\begin{document}

\title{Size reduction of complex networks preserving modularity}

\author{A Arenas\footnote{Author to whom any correspondence should be addressed},
J Duch, A Fern\'andez and S G\'omez}

\address{Departament d'Enginyeria Inform\`{a}tica i Matem\`{a}tiques, Universitat Rovira i Virgili, Avinguda dels Pa\"{\i}sos Catalans 26, 43007 Tarragona, Spain}
\eads{\mailto{alexandre.arenas@urv.cat}, \mailto{jordi.duch@urv.cat}, \mailto{alberto.fernandez@urv.cat} and \mailto{sergio.gomez@urv.cat}}

\begin{abstract}
The ubiquity of modular structure in real-world complex networks is being the focus of attention in many trials to understand the interplay between network topology and functionality. The best approaches to the identification of modular structure are based on the optimization of a quality function known as $modularity$. However this optimization is a hard task provided that the computational complexity of the problem is in the NP-hard class. Here we propose an exact method for reducing the size of weighted (directed and undirected) complex networks while maintaining invariant its modularity. This size reduction allows the heuristic algorithms that optimize modularity for a better exploration of the modularity landscape. We compare the modularity obtained in several real complex-networks by using the Extremal Optimization algorithm, before and after the size reduction, showing the improvement obtained. We speculate that the proposed analytical size reduction could be extended to an exact coarse graining of the network in the scope of real-space renormalization.  
\end{abstract}

\pacno{89.75}
\submitto{\NJP}
\maketitle

\section{Introduction}
The study of the community structure in complex networks is becoming a classical subject in the area because several aspects of the problem are both challenging and interesting. The challenge comes from the difficulty for unveiling the best partition of the network in terms of communities, in the sense of groups of nodes that are more intraconnected rather than interconnected between them \cite{firstnewman}. The interest comes from the fact that this level of description could help to elucidate an organization of the network prescribed by functionalities \cite{amaral,rogernphys}, and also because it resembles the coarse graining process in statistical physics to describe systems at the mesoscale.

The most successful solutions to the community detection problem, in terms of accuracy and computational cost required, are those based in the optimization of a quality function called {\em modularity} proposed by Newman \cite{newgirvan} that allows the comparison of different partitioning of the network. Given a network partitioned into communities, being $C_i$ the community to which node $i$ is assigned, the mathematical definition of modularity is expressed in terms of the weighted adjacency matrix $w_{ij}$, that represents the value of the weight in the link between $i$ and $j$ ($0$ if no link exists), and the strengths
$\ds w_i=\sum_j w_{ij}$ as \cite{newanaly}
\beq
  Q = \frac{1}{2w} \sum_i\sum_j \left(
        w_{ij} - \frac{w_i w_j}{2w}
      \right) \delta(C_i,C_j) \,,
  \label{QW}
\eeq
where the Kronecker delta function $\delta(C_i,C_j)$ takes the values, 1 if nodes $i$ and $j$ are into the same community, 0 otherwise, and the total strength
$\ds 2w=\sum_i w_i =\sum_i \sum_j w_{ij}$.

The modularity of a given partition is then
the probability of having edges falling within groups in the network minus the expected probability in an equivalent (null case) network with the same number of nodes, and edges placed at random preserving the nodes' strength. The larger the value of modularity the best the partitioning is, because more deviates from the null case. Several authors have attacked the problem proposing different optimization heuristics \cite{ newfast, clauset, duch, rogernat, pujol, newspect}  since the number of different partitions are equal to the Bell \cite{bell} or exponential numbers, which grow at least exponentially in the number of nodes $N$. Indeed, optimization of modularity is a NP-hard (Non-deterministic Polynomial-time hard) problem \cite{brandes}. 


The definition of modularity can be also extended, preserving its semantics in terms of probability, to the scenario of weighted directed networks as follows:
\beq
  Q = \frac{1}{2w}\sum_i\sum_j \left(
        w_{ij}-\frac{w_i^{\mbox{\sz out}} w_j^{\mbox{\sz in}}}{2w}
      \right)\delta(C_i,C_j)\,,
\label{QWD}
\eeq
where $w_i^{\mbox{\sz out}}$ and $w_j^{\mbox{\sz in}}$ are respectively the output and input strengths of nodes $i$ and $j$
\begin{eqnarray}
  w_i^{\mbox{\sz out}} & = & \sum_j w_{ij} \,, \\
  w_j^{\mbox{\sz in}}  & = & \sum_i w_{ij} \,,
\end{eqnarray}
and the total strength is
\beq
  2w = \sum_i w_i^{\mbox{\sz out}}
     = \sum_j w_j^{\mbox{\sz in}}
     = \sum_i\sum_j w_{ij} \,.
\eeq
The input and output strengths are equal
($w_i = {w_i}^{\mbox{\sz out}} = {w_i}^{\mbox{\sz in}}$)
if the network is undirected, thus recovering the standard definition of strength. Furthermore, if the network is unweighted and undirected, $w_i$ represents the degree of the $i$-th node, i.e.\ the number of edges attached to it, and $w$ is the total number of links of the network.

The challenge of optimizing the modularity has deserved many efforts from the scientific community in the recent years. Provided the problem is NP-hard, only optimization heuristics have been shown to be competent in finding sub-optimal solutions of $Q$ in feasible computational time. Nevertheless, when facing the decomposition in communities of very large networks, optimality is usually sacrificed in favor of computational time. 

Our goal here is to demonstrate that it is possible to reduce the size of complex networks while preserving the value of modularity, independently on the partition under consideration. The systematic use of this reduction allows for a more exhaustive search of the partitions' space that usually ends in improved values of modularity compared to those obtained without using this size reduction. The paper is organized as follows: In the next section we present the basics for the size reduction process. After that, we provide analytic proofs for specific reductions. Finally we exploit the reduction process based on the mentioned properties, and compare the modularity results with those obtained without size reduction in several real networks, using the Extremal Optimization heuristics \cite{duch}. 

\section{Size reduction preserving modularity}
\subsection{Reduced graph}
Let $G$ be a weighted complex network of size $N$, with weights
$w_{ij}\ge0,\ i,j\in\{1,\ldots,N\}$. If the network is unweighted, the weights matrix becomes the usual connectivity matrix, with values $1$ for connected pairs of nodes, zero otherwise. We will assume that the network may be directed, i.e.\ represented by a non symmetric weights' matrix.

Any grouping of the $N$ nodes of the complex network $G$ in $N'$ parts may be represented by a surjective function
$R:\{1,\ldots,N\} \longrightarrow \{1,\ldots,N'\}$
which assigns a group index $R_i\equiv R(i)$ to every $i$-th node in $G$. 
The {\em reduced network} $G'$ in which each of these groups is replaced by a single node may be easily defined in the following way: the weight $w'_{rs}$ between the nodes which represent groups $r$ and $s$ is the sum of all the weights connecting vertices in these groups,
\beq
  w'_{rs} = \sum_i\sum_j w_{ij}\delta(R_i,r)\delta(R_j,s)\,,\ \
            r,s\in\{1,\ldots,N'\}
\eeq
where the sums run over all the $N$ nodes of $G$. For unweighted networks the value of $w'_{rs}$ is just the number of arcs from the first to the second group of nodes. It must be emphasized that a node $r$ of the reduced network $G'$ acquires a {\em self-loop} if $w'_{rr}\neq 0$, which summarizes the internal connectivity of the nodes of $G$ forming this group.

The input and output strengths of the reduced network $G'$ are
\beq
  {w'_r}^{\mbox{\sz out}} = \sum_s w'_{rs}
                = \sum_i\sum_j w_{ij}\delta(R_i,r)\sum_s\delta(R_j,s)
                = \sum_i {w_i}^{\mbox{\sz out}} \delta(R_i,r)\,,
\eeq
\beq
  {w'_s}^{\mbox{\sz in}} = \sum_r w'_{rs}
                = \sum_j\sum_i w_{ij}\delta(R_j,s)\sum_r\delta(R_i,r)
                = \sum_j {w_j}^{\mbox{\sz in}} \delta(R_j,s)\,,
\eeq
and its total strength $2w'$ is equal to the total strength $2w$ of the original network
\beq
  2w' = \sum_r {w'_r}^{\mbox{\sz out}} = \sum_s {w'_s}^{\mbox{\sz in}}
      = \sum_i {w_i}^{\mbox{\sz out}} = \sum_j {w_j}^{\mbox{\sz in}} = 2w\,.
\eeq

\subsection{Modularity preservation}
The main property of the reduced network is the preservation of modularity (\ref{QW}) or (\ref{QWD}), i.e.\ the modularity of any partition of the reduced graph is equal to the modularity of its corresponding partition of the original network.

More precisely, let
$C':\{1,\ldots,N'\} \longrightarrow \{1,\ldots,M\}$
be a partition in $M$ clusters of the reduced network $G'$. Its corresponding partition
$C:\{1,\ldots,N\} \longrightarrow \{1,\ldots,M\}$
of the original graph is given by the composition of the reducing function $R$ with the partition $C'$, i.e.\ $C=C'\circ R$. Therefore, the statement of the previous paragraph becomes
\beq
  Q'(C') = Q(C)\,.
\eeq
The proof is straightforward:
\begin{eqnarray}
  Q'(C') & = & \frac{1}{2w'}\sum_r\sum_s \left( w'_{rs}
               - \frac{{w'_r}^{\mbox{\sz out}} {w'_s}^{\mbox{\sz in}}}{2w'}
               \right) \delta(C'_r,C'_s) \nn
         & = & \frac{1}{2w}\sum_r\sum_s \left(
               \sum_i\sum_j w_{ij}\delta(R_i,r)\delta(R_j,s)\right. \nn
         &   & \mbox{} - \left.\frac{1}{2w}
                 \sum_i {w_i}^{\mbox{\sz out}}\delta(R_i,r)
                 \sum_j {w_j}^{\mbox{\sz in}}\delta(R_j,s)
               \right) \delta(C'_r,C'_s) \nn
         & = & \frac{1}{2w}\sum_i\sum_j \left( w_{ij}
               - \frac{{w_i}^{\mbox{\sz out}} {w_j}^{\mbox{\sz in}}}{2w}
               \right) \sum_r\sum_s
               \delta(R_i,r)\delta(R_j,s)\delta(C'_r,C'_s)\nn
         & = & \frac{1}{2w}\sum_i\sum_j \left( w_{ij}
               - \frac{{w_i}^{\mbox{\sz out}} {w_j}^{\mbox{\sz in}}}{2w}
               \right) \delta(C'_{R_i},C'_{R_j})\nn
         & = & \frac{1}{2w}\sum_i\sum_j \left( w_{ij}
               - \frac{{w_i}^{\mbox{\sz out}} {w_j}^{\mbox{\sz in}}}{2w}
               \right) \delta(C_i,C_j) \nn
         & = & Q(C)
\end{eqnarray}

We have found a relevant property of modularity namely that those nodes forming a community in the optimal partition can be represented by a unique node in the reduced network. Each node in the reduced network summarizes the information necessary for the calculation of modularity in its self-loop (that accounts for the intraconnectivity of the community) and its arcs (that account for the total strengths with the rest of the network). The question now is: how to determine which nodes will belong to the same community in the optimal partition, before this partition is obtained? The answer will provide with a size reduction method in complex networks preserving modularity.

\section{Analytic reductions}
Here we give the proof for certain possible analytic size reductions of weighted networks, undirected and directed.

\subsection{Reductions for undirected networks}
The modularity of an undirected network may be written as
\beq
  Q =\sum_i q_i\,,
\eeq
where
\beq
  q_i = \frac{1}{2w}\sum_j \left(
          w_{ij} - \frac{w_i w_j}{2w}
        \right) \delta(C_i,C_j)
\eeq
is the contribution to modularity of the $i$-th node. If we allow this node to change of community, the value of $C_i$ becomes a parameter, and therefore it is useful to define
\beq
  q_{i,r} = \frac{1}{2w}\sum_j \left(
              w_{ij} - \frac{w_i w_j}{2w}
            \right) \delta(C_j,r)\,,
            \ \ \  q_i = q_{i,C_i}\,,
\eeq
which accounts for the contribution of the $i$-th node to modularity if it were in community $r$. The separation of the self-loop term, which does not depend on which community node $i$ belongs to, yields to the definition of
\beq
  {\tilde{q}}_{i,r} = \frac{1}{2w} \sum_{j(\neq i)} \left(
                        w_{ij} - \frac{w_i w_j}{2w}
                      \right) \delta(C_j,r)\,,
            \ \ \  {\tilde{q}}_i = {\tilde{q}}_{i,C_i}
\eeq
and
\beq
  \tilde{Q} = \sum_i {\tilde{q}}_i
            = \frac{1}{2w} \sum_i \sum_{j(\neq i)} \left(
                w_{ij} - \frac{w_i w_j}{2w}
              \right) \delta(C_j,r)\,,
\eeq
satisfying
\beq
  q_{i,r} = {\tilde{q}}_{i,r} + \frac{1}{2w} \left(
              w_{ii} - \frac{w_i^2}{2w}
            \right)
\eeq
and
\beq
  Q = \tilde{Q} + \frac{1}{2w} \sum_i \left(
        w_{ii} - \frac{w_i^2}{2w}
      \right)\,.
\eeq

The role of these individual node contributions to modularity becomes evident in the expression of the change of modularity when node $i$ goes from community $r$ to community $s$:
\beq
  \Delta Q = 2 ( {\tilde{q}}_{i,s} - {\tilde{q}}_{i,r})\,.
\eeq
As a particular case, a node that forms its own community, i.e.\ an isolated node $i$, which moves to any community $s$ produces a change in modularity
\beq \label{deltaqiso}
  \Delta Q = 2 {\tilde{q}}_{i,s}\,.
\eeq
Therefore, if there exists a community $s$ for which ${\tilde{q}}_{i,s} > 0$, node $i$ cannot be isolated in the partition of optimal modularity. This existence is easily proved by considering the sum of ${\tilde{q}}_{i,r}$ for all communities:
\begin{eqnarray} 
  \sum_r {\tilde{q}}_{i,r} & = &
    \frac{1}{2w} \sum_{j(\neq i)} \left(
      w_{ij} - \frac{w_i w_j}{2w}
    \right) \sum_r \delta(C_j,r) \nn
  & = &
    \frac{1}{2w} \sum_{j(\neq i)} \left(
      w_{ij} - \frac{w_i w_j}{2w}
    \right) \nn
  & = & \label{sumrqtilde}
    - \frac{1}{2w} \left( w_{ii} - \frac{w_i^2}{2w} \right)\,.
\end{eqnarray}
where we have made use of the definitions of strength $w_i$ and total strength $2w$ for the simplification of the expression. Thus,
\beq
  \mbox{if } w_{ii} \leq\ \frac{w_i^2}{2w} \ \ \Rightarrow \ \
    \sum_r {\tilde{q}}_{i,r} \geq 0 \ \ \Rightarrow \ \
    \exists s: {\tilde{q}}_{i,s} \geq 0\,,
\eeq
completing the proof that there are no isolated nodes in the configuration which maximizes modularity, unless they have a big enough self-loop \footnote{Note that some authors \cite{massen} have used the fact that no isolated nodes are obtained at the partition of maximum modularity to reduce the network size, simply by obviating the existence of these nodes. This approach clearly fails to reproduce the same modularity of the original network and provides misleading results, it should be avoided.}.

Now, it remains the problem of the determination of an acquaintance (node $j$) of node $i$ in its optimal community, in order to group them ($R_i = R_j$) in a single equivalent node with a self-loop, as explained above. If we know that nodes $i$ and $j$ share the same community at maximum modularity, the reduced network will be equivalent to the original one as regards modularity: no information lost, and a smaller size. Taking into account that the sign of the ${\tilde{q}}_{i,r}$ can only be positive if there is a link between node $i$ and another node in community $r$, the only candidates to be the right acquaintance of any node are its neighbors in the network. 

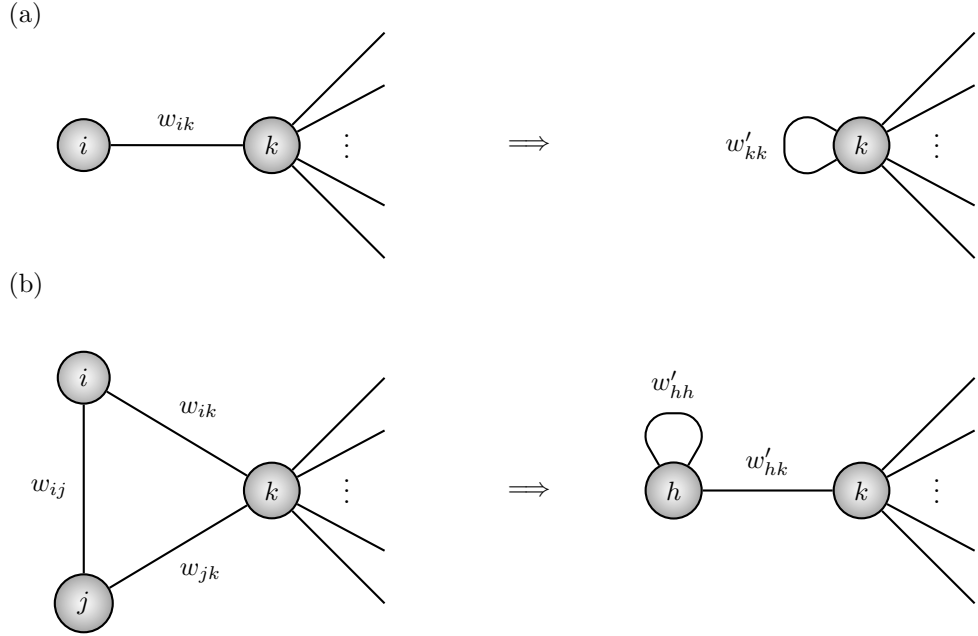
\begin{figure}[t]
\begin{indented}
\item[]
\begin{tabular}{@{}lcr}
  (a) \\
  \begin{pspicture}(0,1)(6,4)
    \psset{framesep=0.2,nodesep=0}
    \psset{fillstyle=ccslope,slopebegin=white,slopeend=gray,sloperadius=0.5}
    \cnodeput(1,2.5){a}{$i$}
    \cnodeput(3.5,2.5){c}{$k$}
    \psset{fillstyle=none}
    \pnode(5,4){p}
    \pnode(5,3.3){q}
    \cnodeput*(4.5,2.6){r}{$\vdots$}
    \pnode(5,1.7){s}
    \pnode(5,1){t}
    \ncline{-}{a}{c}\Aput{$w_{ik}$}
    \ncline{-}{c}{p}
    \ncline{-}{c}{q}
    \ncline{-}{c}{s}
    \ncline{-}{c}{t}
  \end{pspicture}
  &
  \begin{pspicture}(0,1)(1,4)
    \cput*(0.5,2.5){$\Longrightarrow$}
  \end{pspicture}
  &
  \begin{pspicture}(0,1)(6,4)
    \psset{framesep=0.2,nodesep=0}
    \psset{fillstyle=ccslope,slopebegin=white,slopeend=gray,sloperadius=0.5}
    \cnodeput(3.5,2.5){c}{$k$}
    \psset{fillstyle=none}
    \pnode(5,4){p}
    \pnode(5,3.3){q}
    \cnodeput*(4.5,2.6){r}{$\vdots$}
    \pnode(5,1.7){s}
    \pnode(5,1){t}
    \ncdiag[arm=0.8,angleA=150,angleB=210,linearc=0.3]{c}{c}\Bput{$w'_{kk}$}
    \ncline{-}{c}{p}
    \ncline{-}{c}{q}
    \ncline{-}{c}{s}
    \ncline{-}{c}{t}
  \end{pspicture}
  \\
  (b) \\
  \begin{pspicture}(0,0)(6,5)
    \psset{framesep=0.2,nodesep=0}
    \psset{fillstyle=ccslope,slopebegin=white,slopeend=gray,sloperadius=0.5}
    \cnodeput(1,4){a}{$i$}
    \cnodeput(1,1){b}{$j$}
    \cnodeput(3.5,2.5){c}{$k$}
    \psset{fillstyle=none}
    \pnode(5,4){p}
    \pnode(5,3.3){q}
    \cnodeput*(4.5,2.6){r}{$\vdots$}
    \pnode(5,1.7){s}
    \pnode(5,1){t}
    \ncline{-}{a}{b}\Bput{$w_{ij}$}
    \ncline{-}{a}{c}\Aput{$w_{ik}$}
    \ncline{-}{b}{c}\Bput{$w_{jk}$}
    \ncline{-}{c}{p}
    \ncline{-}{c}{q}
    \ncline{-}{c}{s}
    \ncline{-}{c}{t}
  \end{pspicture}
  &
  \begin{pspicture}(0,0)(1,5)
    \cput*(0.5,2.5){$\Longrightarrow$}
  \end{pspicture}
  &
  \begin{pspicture}(0,0)(6,5)
    \psset{framesep=0.2,nodesep=0}
    \psset{fillstyle=ccslope,slopebegin=white,slopeend=gray,sloperadius=0.5}
    \cnodeput(1,2.5){a}{$h$}
    \cnodeput(3.5,2.5){c}{$k$}
    \psset{fillstyle=none}
    \pnode(5,4){p}
    \pnode(5,3.3){q}
    \cnodeput*(4.5,2.6){r}{$\vdots$}
    \pnode(5,1.7){s}
    \pnode(5,1){t}
    \ncline{-}{a}{c}\Aput{$w'_{hk}$}
    \ncdiag[arm=0.8,angleA=60,angleB=120,linearc=0.3]{a}{a}\Bput{$w'_{hh}$}
    \ncline{-}{c}{p}
    \ncline{-}{c}{q}
    \ncline{-}{c}{s}
    \ncline{-}{c}{t}
  \end{pspicture}
\end{tabular}
\end{indented}
  \caption{Analytic reductions for undirected networks.
    In (a) example of a {\em hair} reduction,
    (b) example of a {\em triangular hair} reduction (see text for details). The widespread case of unweighted networks, all weights equal to 1, implies that in the reduction (a), $w'_{kk}=2$, and in the reduction (b), $w'_{hh}=2$ and $w'_{hk}=2$.}
  \label{fig1}
\end{figure}

The simplest particular cases are {\em hairs}, i.e.\ nodes connected to the network with only one link. Hence, a hair can be analytically grouped with its neighbor $k$ if 
\beq
  w_{ii}\leq\frac{w_i^2}{2w}\,,
\eeq
producing a self-loop for node $k$ of value
\beq
  w'_{kk}=w_{ii}+2w_{ik}\,.
\eeq
When node $i$ has no self-loop ($w_{ii}=0$) this condition is always fulfilled, see figure~\ref{fig1}a.
  
Another solvable structure is the {\em triangular hair}, in which two nodes $i$ and $j$ have only one link connecting them, two more links from $i$ and $j$ to a third node $k$, and possibly self-loops. In this case, if 
\beq
  w_{ii}\leq\frac{w_i^2}{2w}\ \mbox{ and } w_{jj}\leq\frac{w_j^2}{2w}
\eeq
nodes $i$ and $j$ share the same community in the optimal partition and therefore may be grouped as a single node $h$. Moreover, the resulting structure becomes a simple hair, which can be grouped with node $k$ if
\beq
  w'_{hh}\leq\frac{w_h^{'2}}{2w'}\, 
\eeq
where 
\begin{eqnarray}
  w'_{hh} & = w_{ii}+2w_{ij}+w_{jj}\,, \nn
  w'_{hk} & = w_{ik}+w_{jk}\,, \nn
  w'_h & = w_i+w_j = w'_{hh}+w'_{hk}\,.
\end{eqnarray}
In the particular case of nodes $i$ and $j$ without self-loops ($w_{ii}=w_{jj}=0$), the triangular hair can always be reduced to a single hair with a self-loop $w'_{hh}=2w_{ij}$, see figure~\ref{fig1}b.

\subsection{Reductions for directed networks}
The treatment of directed networks requires the distinction between the nodes' output and input contributions to modularity:
\beq
  Q = \sum_i q_i^{\mbox{\sz out}} = \sum_j q_j^{\mbox{\sz in}}\,,
\eeq
where
\beq
  q_{i,r}^{\mbox{\sz out}} = \frac{1}{2w}\sum_j \left(
      w_{ij}-\frac{w_i^{\mbox{\sz out}} w_j^{\mbox{\sz in}}}{2w}
    \right)\delta(C_j,r)\,,
    \ \ \  q_i^{\mbox{\sz out}} = q_{i,C_i}^{\mbox{\sz out}}\,,
\eeq
\beq
  q_{j,r}^{\mbox{\sz in}} = \frac{1}{2w}\sum_i \left(
      w_{ij}-\frac{w_i^{\mbox{\sz out}} w_j^{\mbox{\sz in}}}{2w}
    \right)\delta(C_i,r)\,,
    \ \ \  q_j^{\mbox{\sz in}} = q_{j,C_j}^{\mbox{\sz in}}\,.
\eeq

The separation of the self-loop term follows the same pattern than for undirected networks:
\beq
  {\tilde{q}}_{i,r}^{\mbox{\sz out}} = \frac{1}{2w}\sum_{j(\neq i)} \left(
      w_{ij}-\frac{w_i^{\mbox{\sz out}} w_j^{\mbox{\sz in}}}{2w}
    \right)\delta(C_j,r)\,,
    \ \ \  {\tilde{q}}_i^{\mbox{\sz out}} = {\tilde{q}}_{i,C_i}^{\mbox{\sz out}}\,,
\eeq
\beq
  {\tilde{q}}_{j,r}^{\mbox{\sz in}} = \frac{1}{2w}\sum_{i(\neq j)} \left(
      w_{ij}-\frac{w_i^{\mbox{\sz out}} w_j^{\mbox{\sz in}}}{2w}
    \right)\delta(C_i,r)\,,
    \ \ \  {\tilde{q}}_j^{\mbox{\sz in}} = {\tilde{q}}_{j,C_j}^{\mbox{\sz in}}\,,
\eeq
and
\beq
  \tilde{Q} = \sum_i \tilde{q}_i^{\mbox{\sz out}}
            = \sum_j \tilde{q}_j^{\mbox{\sz in}}\,,
\eeq
satisfying
\beq
  q_{i,r}^{\mbox{\sz out}} = {\tilde{q}}_{i,r}^{\mbox{\sz out}}
    + \frac{1}{2w} \left(
      w_{ii} - \frac{w_i^{\mbox{\sz out}} w_i^{\mbox{\sz in}}}{2w}
    \right)\,,
\eeq
\beq
  q_{j,r}^{\mbox{\sz in}} = {\tilde{q}}_{j,r}^{\mbox{\sz in}}
    + \frac{1}{2w} \left(
      w_{jj} - \frac{w_j^{\mbox{\sz out}} w_j^{\mbox{\sz in}}}{2w}
    \right)\,,
\eeq
and
\beq
  Q = \tilde{Q} + \frac{1}{2w} \sum_i \left(
        w_{ii} -  \frac{w_i^{\mbox{\sz out}} w_i^{\mbox{\sz in}}}{2w}
      \right)\,.
\eeq

With these definitions at hand, the change of modularity when node $i$ goes from community $r$ to community $s$ becomes
\beq
  \Delta Q =
    ({\tilde{q}}_{i,s}^{\mbox{\sz out}} + {\tilde{q}}_{i,s}^{\mbox{\sz in}})
    - ({\tilde{q}}_{i,r}^{\mbox{\sz out}} + {\tilde{q}}_{i,r}^{\mbox{\sz in}})\,,
\eeq
and the change when an isolated node $i$ moves to any community $s$ is
\beq \label{deltaqisodir}
  \Delta Q =
    {\tilde{q}}_{i,s}^{\mbox{\sz out}} + {\tilde{q}}_{i,s}^{\mbox{\sz in}}\,.
\eeq

The first difference between directed and undirected networks comes from the fact that we cannot prove this time the inexistence of isolated nodes in the partition of optimal modularity. The previous argumentation was based on the use of (\ref{sumrqtilde}), which now splits in two relationships:
\beq
  \sum_r {\tilde{q}}_{i,r}^{\mbox{\sz out}} = - \frac{1}{2w} \left(
    w_{ii} - \frac{w_i^{\mbox{\sz out}} w_i^{\mbox{\sz in}}}{2w}
  \right) \,,
\eeq
\beq 
  \sum_r {\tilde{q}}_{j,r}^{\mbox{\sz in}} = - \frac{1}{2w} \left(
    w_{jj} - \frac{w_j^{\mbox{\sz out}} w_j^{\mbox{\sz in}}}{2w}
  \right) \,.
\eeq
The next step is the same:
\beq \label{sumqtirout}
  \mbox{if } w_{ii} \leq\ \frac{w_i^{\mbox{\sz out}} w_i^{\mbox{\sz in}}}{2w}
    \ \ \Rightarrow \ \
    \sum_r {\tilde{q}}_{i,r}^{\mbox{\sz out}} \geq 0 \ \ \Rightarrow \ \
    \exists s_1: {\tilde{q}}_{i,s_1}^{\mbox{\sz out}} \geq 0\,,
\eeq
\beq \label{sumqtirin}
  \mbox{if } w_{ii} \leq\ \frac{w_i^{\mbox{\sz out}} w_i^{\mbox{\sz in}}}{2w}
    \ \ \Rightarrow \ \
    \sum_r {\tilde{q}}_{i,r}^{\mbox{\sz in}} \geq 0 \ \ \Rightarrow \ \
    \exists s_2: {\tilde{q}}_{i,s_2}^{\mbox{\sz in}} \geq 0\,.
\eeq
Since communities $s_1$ and $s_2$ need not be the same, the change of modularity (\ref{deltaqisodir}) is not warranted to be positive, and thus isolated nodes are possible in the partition which maximizes modularity.

Nevertheless, there exist three kinds of nodes for which we can prove they cannot be isolated in the optimal partition, provided their self-loops are not too large: hairs, {\em sinks} (nodes with only input links) and {\em sources} (nodes with only output links).

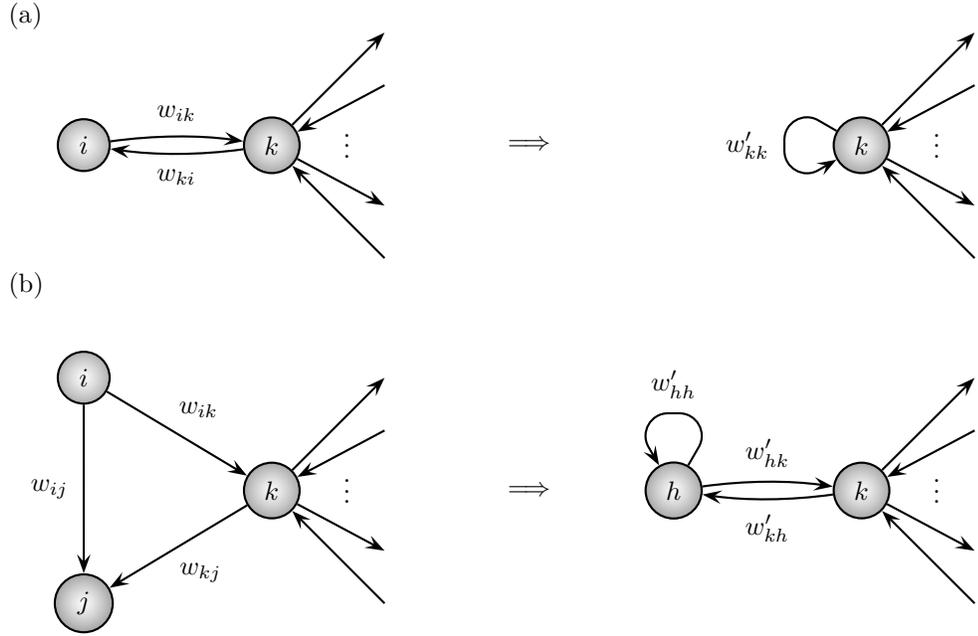
\begin{figure}[t]
\begin{indented}
\item[]
\begin{tabular}{@{}lcr}
  (a) \\
  \begin{pspicture}(0,1)(6,4)
    \psset{framesep=0.2,nodesep=0,arrowsize=2pt 4}
    \psset{fillstyle=ccslope,slopebegin=white,slopeend=gray,sloperadius=0.5}
    \cnodeput(1,2.5){a}{$i$}
    \cnodeput(3.5,2.5){c}{$k$}
    \psset{fillstyle=none}
    \pnode(5,4){p}
    \pnode(5,3.3){q}
    \cnodeput*(4.5,2.6){r}{$\vdots$}
    \pnode(5,1.7){s}
    \pnode(5,1){t}
    \ncarc{->}{a}{c}\Aput{$w_{ik}$}
    \ncarc{->}{c}{a}\Aput{$w_{ki}$}
    \ncline{->}{c}{p}
    \ncline{<-}{c}{q}
    \ncline{->}{c}{s}
    \ncline{<-}{c}{t}
  \end{pspicture}
  &
  \begin{pspicture}(0,1)(1,4)
    \cput*(0.5,2.5){$\Longrightarrow$}
  \end{pspicture}
  &
  \begin{pspicture}(0,1)(6,4)
    \psset{framesep=0.2,nodesep=0,arrowsize=2pt 4}
    \psset{fillstyle=ccslope,slopebegin=white,slopeend=gray,sloperadius=0.5}
    \cnodeput(3.5,2.5){c}{$k$}
    \psset{fillstyle=none}
    \pnode(5,4){p}
    \pnode(5,3.3){q}
    \cnodeput*(4.5,2.6){r}{$\vdots$}
    \pnode(5,1.7){s}
    \pnode(5,1){t}
    \ncdiag[arm=0.8,angleA=150,angleB=210,linearc=0.3]{->}{c}{c}\Bput{$w'_{kk}$}
    \ncline{->}{c}{p}
    \ncline{<-}{c}{q}
    \ncline{->}{c}{s}
    \ncline{<-}{c}{t}
  \end{pspicture}
  \\
  (b) \\
  \begin{pspicture}(0,0)(6,5)
    \psset{framesep=0.2,nodesep=0,arrowsize=2pt 4}
    \psset{fillstyle=ccslope,slopebegin=white,slopeend=gray,sloperadius=0.5}
    \cnodeput(1,4){a}{$i$}
    \cnodeput(1,1){b}{$j$}
    \cnodeput(3.5,2.5){c}{$k$}
    \psset{fillstyle=none}
    \pnode(5,4){p}
    \pnode(5,3.3){q}
    \cnodeput*(4.5,2.6){r}{$\vdots$}
    \pnode(5,1.7){s}
    \pnode(5,1){t}
    \ncline{->}{a}{b}\Bput{$w_{ij}$}
    \ncline{->}{a}{c}\Aput{$w_{ik}$}
    \ncline{<-}{b}{c}\Bput{$w_{kj}$}
    \ncline{->}{c}{p}
    \ncline{<-}{c}{q}
    \ncline{->}{c}{s}
    \ncline{<-}{c}{t}
  \end{pspicture}
  &
  \begin{pspicture}(0,0)(1,5)
    \cput*(0.5,2.5){$\Longrightarrow$}
  \end{pspicture}
  &
  \begin{pspicture}(0,0)(6,5)
    \psset{framesep=0.2,nodesep=0,arrowsize=2pt 4}
    \psset{fillstyle=ccslope,slopebegin=white,slopeend=gray,sloperadius=0.5}
    \cnodeput(1,2.5){a}{$h$}
    \cnodeput(3.5,2.5){c}{$k$}
    \psset{fillstyle=none}
    \pnode(5,4){p}
    \pnode(5,3.3){q}
    \cnodeput*(4.5,2.6){r}{$\vdots$}
    \pnode(5,1.7){s}
    \pnode(5,1){t}
    \ncarc{->}{a}{c}\Aput{$w'_{hk}$}
    \ncarc{->}{c}{a}\Aput{$w'_{kh}$}
    \ncdiag[arm=0.8,angleA=60,angleB=120,linearc=0.3]{->}{a}{a}\Bput{$w'_{hh}$}
    \ncline{->}{c}{p}
    \ncline{<-}{c}{q}
    \ncline{->}{c}{s}
    \ncline{<-}{c}{t}
  \end{pspicture}
\end{tabular}
\end{indented}
  \caption{Analytic reductions for directed networks.
    In (a) example of a {\em hair} reduction,
    (b) example of a {\em triangular hair} reduction (see text for details)}
  \label{fig2}
\end{figure}

Directed hairs, i.e.\ nodes connected only to another node, either through an input, an output, or both links, necessarily have $s_1=s_2$. Therefore, it is save to group them in the same way as undirected hairs if
\beq
  w_{ii} \leq\ \frac{w_i^{\mbox{\sz out}} w_i^{\mbox{\sz in}}}{2w}\,.
\eeq
In particular, this condition is always fulfilled if the hair has no self-loop ($w_{ii}=0$), see figure \ref{fig2}a. Whenever the self-loop is present, both input and output links are needed to counterbalance it. The resulting self-loop $w'_{kk}$ of the grouped node has value
\beq
  w'_{kk} = w_{ii} + w_{ik} + w_{ki}\,.
\eeq

Sink nodes $i$ are characterized by null output strengths, $w_i^{\mbox{\sz out}}=0$, which imply ${\tilde{q}}_{i,r}^{\mbox{\sz out}}=0$ for all communities $r$. Thus, the change of modularity (\ref{deltaqisodir}) only depends on the value of ${\tilde{q}}_{i,s}^{\mbox{\sz in}}$, and (\ref{sumqtirin}) tells us that they can always be grouped with an increase of modularity. The same property applies to sources, which are defined as nodes with null input strengths, $w_i^{\mbox{\sz in}}=0$. Note that sinks and sources cannot have self-loops, since this would be in contradiction with their null output and input strengths respectively.

A triangular hair formed by a source node $i$ and a sink node $j$ behaves exactly as the undirected triangular hair, being possible to group them in a single node $h$ with a self-loop, see figure \ref{fig2}b, where 
\begin{eqnarray}
  w'_{hh} & = w_{ij}\,, \nn
  w'_{hk} & = w_{ik}\,, \nn
  w'_{kh} & = w_{kj}\,.
\end{eqnarray}

\section{Results and discussion}
The above proofs allow us to face the problem of size reduction in complex networks into a firm basis. In particular, this size reduction preserving modularity ensures that the structural mesoscale found by maximizing modularity will be invariant under these transformations. The natural question at this point is: what is the percentage in size reduction of networks using the previous rules? To answer this question it is mandatory to have an estimation on the number of hairs, and triangular hairs, we might expect in complex networks. In real networks this calculation can be performed by direct enumeration, however an estimation can be made in terms of general grounds about the degree distribution $P(k)$. 
\begin{table}[ht]
\caption{\label{res}Results for the optimal partition obtained using EO algorithm \cite{duch} for several real networks before and after applying the size reduction. We present the number of nodes, modularity, number of communities and speed-up of the algorithm after reduction.}

\begin{indented}
\item[]
\begin{tabular}{@{}lrrrr}
\br
Network & $N$ & $Q$ & \# communities & speed-up \\
\mr
Zachary & 34 & 0.419790 & 4 & -- \\
Zachary-reduced & 33 & 0.419790 & 4 & 1.00 \\
Jazz & 198 & 0.444469 & 4 & --\\
Jazz-reduced & 193 & 0.445144 & 4 & 1.00 \\
E-mail & 1133 & 0.580070 & 10 & --   \\
E-mail-reduced & 981 & 0.581425 & 10 & 1.17   \\
Airports-U & 3618 & 0.706704 & 25 & -- \\
Airports-U-reduced & 2763 & 0.707076 & 24 & 1.68  \\
Airports-WU & 3618 & 0.649268 & 29 & --  \\
Airports-WU-reduced & 2763 & 0.649337 & 29 &  1.68 \\
Airports-WD & 3618 & 0.649189 & 34 & --  \\
Airports-WD-reduced & 2880 & 0.649286 & 30 & 1.53  \\
PGP & 10680 & 0.876883 & 118 & --  \\
PGP-reduced & 6277 & 0.880244 & 101 & 4.27  \\
AS(2001) & 11174 & 0.619048 & 25 & --  \\
AS(2001)-reduced & 7386 & 0.628004 & 31 & 2.41  \\
AS(2006) & 22963 & 0.645942 & 25 & -- \\
AS(2006)-reduced & 15118 & 0.658198 & 45 & 2.39  \\
\br
\end{tabular}
\end{indented}
\end{table}

Here we provide some rough estimates for the most widespread degree distributions in natural and artificial networks: scale-free and exponential. For scale-free networks it is usually assumed a $P(k)=\alpha k^{-\gamma}$, with $\gamma\in[2,3]$ for most of the real scale-free complex networks. The normalization condition provides with the value of $\alpha$. As a first approximation, neglecting the structural cut-off of the network, we can write 
\beq
\alpha\sum_{k=1}^{\infty} k^{-\gamma}=\alpha \zeta(\gamma)=1
\eeq 
\noindent where $\zeta(\gamma)$ is the Dirichlet series representation of the Riemman zeta function. For values of $\gamma\in[2,3]$ we obtain $\alpha\in[1/\zeta(2),1/\zeta(3)]\approx[0.61,0.83]$. That means that, roughly speaking, the number of hairs that corresponds to $P(1)$ is about 83\% of nodes  in a scale-free network with $\gamma=3$ and 61\% when $\gamma=2$, although this value is slightly reduced when considering the cut-offs of the real distributions.
 
An equivalent estimate can be conducted for exponential degree distributions of type $P(k)=\alpha e^{-\beta k}$, with $\beta>0$. In this case, normalization implies that
\beq
\alpha\sum_{k=1}^{\infty}  e^{-\beta k}=\alpha \frac{e^{-\beta}}{1-e^{-\beta}}=1
\eeq
\noindent and then $\alpha=e^{\beta}-1$. The percentage of hairs in this case is $P(1)=1-e^{-\beta}$, that, for example, for plausible values of $\beta\in[0.5,1.5]$ provides a reduction between 40\% and 77\% respectively.

At the light of these estimates, the size reduction process provides with an interesting technique to confront the analysis of community structure in networks by maximizing modularity with a substantial advantage in computational cost without sacrificing any information. We have checked our size reduction process, and posterior optimization of modularity using Extremal Optimization (EO) \cite{duch} 
in several real networks. To enhance the accuracy of the EO algorithm, we perform a last step of optimization consisting in to merge communities whenever modularity is increased, and rearrange the borders (moving the nodes with the lowest modularity values and testing them in the neighbor communities) until all the nodes are better classified and no higher modularities, by moving one node, can be obtained. The results obtained improve those obtained using Spectral optimization \cite{newspect} and simulated annealing \cite{rogernat}. 

The networks analyzed are: the Zachary's karate club network \cite{zachary}, the Jazz musicians network \cite{leon}, the e-mail network of the University Rovira i Virgili \cite{email}, the airports network with data about passenger flights operating in the time period November 1, 2000, to October 31, 2001 compiled by OAG Worldwide (Downers Grove, IL) and analyzed in \cite{airports}, the network of users of the PGP algorithm for secure information transactions \cite{PGP}, and the Internet network at the autonomous system (AS) level as it was in 2001 and 2006 reconstructed from BGP tables posted by the University of Oregon Route Views Project. The results obtained are reported in Table 1. 

We observe that the reduction process allows for a more exhaustive search of the partitions' space as expected. The speed-up of the algorithm after reduction gives an indication of the effectiveness of the process. This is also corroborated by an improvement in modularity. We present in Table 1 the values of modularity for the different networks analyzed up to order $10^{-6}$. In general, the numerical resolution of modularity is up to order $\min_i\{w_{i}\}/2w$, that represents the minimal possible change in the structure of the partitions. It means that every digit in our value of modularity is significant for comparison purposes.

Particularly illustrative is the analysis of the airport network. We have constructed different networks from the raw data, the undirected unweighted network previously used in \cite{airports}, the undirected weighted network (where the weights reflects the number of passengers using the connection in the period of study), and the most realistic case corresponding to the weighted directed network of the airports connections. These networks allowed us to check our techniques (reduction and optimization algorithm) in all the possible scenarios. Note that the results obtained for the weighted directed and undirected networks in terms of modularity are very close, an explanation about this fact that is ubiquitous in the analysis of directed networks can be found in the Appendix.

Summarizing, we have proposed an exact procedure for size reduction in complex networks preserving modularity. The direct consequence of its application is an improvement in computational cost, and then accuracy, of any heuristics designed to optimize modularity. We think that the idea of the exact reduction could be extended to other specific motifs (building blocks) in the network, although its analytical treatment can be further difficult. The reduced network is also an appealing concept to renormalize dynamical processes in complex networks (in the sense of real space renormalization). With this reduction it is plausible to perform a coarse graining of the dynamic interactions between the formed groups, we will explore this connection in a future work.

\section*{Acknowledgements}
We thank L.A.N. Amaral group for providing the airports data set. This work has been supported by the Spanish DGICYT Project FIS2006-13321-C02-02.

\appendix
\section{Relationship between directed and undirected modularities}
Let us suppose that $w_{ij}$ are the weights of a directed weighted network, and that we define its corresponding symmetrized (undirected) network by adding the weights matrix to its transpose:
\beq
  \bar{w}_{ij}=w_{ij}+w_{ji}\,,\ \forall i,j\,.
\eeq
The strengths of this undirected network are
\beq
  \bar{w}_{i}=w_i^{\mbox{\sz out}}+w_i^{\mbox{\sz in}}\,,
\eeq
and the total strength is
\beq
  2\bar{w}=4w\,.
\eeq

The modularity $Q_D$ of the directed network is invariant under transposition of the weights matrix since the input (output) strengths of the transposed network are equal to the output (input) strengths of the original one:
\begin{eqnarray}
  Q_D &=& \frac{1}{2w}\sum_i\sum_j\left(
            w_{ij}-\frac{w_i^{\mbox{\sz out}}w_j^{\mbox{\sz in}}}{2w}
          \right)\delta(C_i,C_j)\nn
      &=& \frac{1}{2w}\sum_j\sum_i\left(
            w_{ji}-\frac{w_j^{\mbox{\sz out}}w_i^{\mbox{\sz in}}}{2w}
          \right)\delta(C_j,C_i)\nn
      &=& \frac{1}{2w}\sum_i\sum_j\left(
            w_{ji}-\frac{w_i^{\mbox{\sz in}}w_j^{\mbox{\sz out}}}{2w}
          \right)\delta(C_i,C_j)\,.
\end{eqnarray}

The relationship between the modularity $Q_D$ of the directed network and the modularity $Q_S$ of its symmetrization is obtained by simple calculations:
\begin{eqnarray}
  Q_S &=& \frac{1}{2\bar{w}}\sum_i\sum_j\left(
            \bar{w}_{ij}-\frac{\bar{w}_{i}\bar{w}_{j}}{2\bar{w}}
          \right)\delta(C_i,C_j)\nn
      &=& \frac{1}{4w}\sum_i\sum_j\left(
            w_{ij}+w_{ji}-\frac{(w_i^{\mbox{\sz out}}+w_i^{\mbox{\sz in}})
            (w_j^{\mbox{\sz out}}+w_j^{\mbox{\sz in}})}{4w}
          \right)\delta(C_i,C_j)\nn
      &=& \frac{1}{4w}\sum_i\sum_j \left[
            \left(
              w_{ij}-\frac{w_i^{\mbox{\sz out}}w_j^{\mbox{\sz in}}}{2w}
            \right)
          +
            \left(
              w_{ji}-\frac{w_i^{\mbox{\sz in}}w_j^{\mbox{\sz out}}}{2w}
            \right)\right]\delta(C_i,C_j)\nn
      & & \mbox{} - \frac{1}{(4w)^2}\sum_i\sum_j
            (w_i^{\mbox{\sz out}}-w_i^{\mbox{\sz in}})
            (w_j^{\mbox{\sz out}}-w_j^{\mbox{\sz in}})
          \delta(C_i,C_j)\nn
      &=& Q_D - \frac{1}{(4w)^2}\sum_i\sum_j
            (w_i^{\mbox{\sz out}}-w_i^{\mbox{\sz in}})
            (w_j^{\mbox{\sz out}}-w_j^{\mbox{\sz in}})
          \delta(C_i,C_j)\,.
\end{eqnarray}
This result can also be expressed as a communities sum:
\beq  Q_S = Q_D - \frac{1}{(4w)^2}\sum_{r} \left(
            \sum_i (w_i^{\mbox{\sz out}}-w_i^{\mbox{\sz in}})\delta(C_i,r)
          \right)^2\,.
\eeq
The contribution of the links to the input and output strengths cancel if they fall within the communities. Therefore, if most links do not cross the boundaries of the communities, it follows that $Q_S \approx Q_D$ even if the network is highly asymmetric.

\end{document}